\documentclass[aps,prb,twocolumn,groupedaddress,10pt,amssymb,amsfonts]{revtex4-1}
\usepackage{amsmath,latexsym,hyperref}
\usepackage{graphicx}

\def\beq{\begin{equation}}
\def\eeq{\end{equation}}
\def\be{\begin{equation}}
\def\ee{\end{equation}}

\renewcommand{\bf}{\mathbf}
\def\d{\partial}

\def\tr{\hbox{tr}\,}

\def\cH{{\cal H}}
\def\rf#1{(\ref{#1})}
\def\rfs#1{Eq.~\rf{#1}}

\begin{document}

\title{ Delocalization of boundary states  in disordered topological insulators}
\author{Andrew M.~Essin$^1$}
\author{Victor~Gurarie$^2$}
\affiliation{$^1$Institute for Quantum Information and Matter and Department of Physics,
California Institute of Technology, Pasadena CA 91125, USA}
\affiliation{$^2$Department of Physics, CB390, University of Colorado, Boulder CO 80309, USA}
\date{\today}

\begin{abstract}
We use the method of bulk-boundary correspondence of topological invariants to show that disordered topological insulators have at least one delocalized state at their boundary at zero energy. Those insulators which do not have chiral (sublattice) symmetry have in addition the whole band of delocalized states at their boundary, with the zero energy state lying in the middle of the band. This result was previously conjectured based on the anticipated properties of the supersymmetric (or replicated) sigma models with WZW-type terms, as well as verified in some cases using numerical simulations and a variety of other arguments. Here we derive this result generally, in arbitrary number of dimensions, and without relying on the description in the language of sigma models. 
\end{abstract}

\maketitle

\section{Introduction}
Topological insulators are noninteracting fermionic systems which are bulk insulators, have gapless excitations at their boundary,  and which are characterized by topological invariants. As free fermion systems, they are described by the Hamiltonian
\be \hat H = \sum_{\alpha \beta} {\cal H}_{\alpha \beta} \hat a^\dagger_{\alpha} \hat a_{\beta}.
\ee
Here $\alpha, \beta$ label points in space, spin and flavor of the fermions. 

A procedure developed by G. Volovik in the 80s \cite{VolovikBook1} to relate the edge states of two dimensional integer Hall systems to the bulk topological invariants in a direct way was recently generalized to all topological insulators \cite{Essin2011}.  In effect, that procedure showed that an edge of a topological insulator is a topological metal, characterized by its own topological invariant whose value must be equal to the value of the bulk invariant. All of this was done in the absence of disorder. 

Here we show that a suitable modification of this formalism extends it to the case when the topological insulators are disordered. It immediately follows
from this formalism that the edge of topological insulators cannot be fully localized by disorder. Furthermore, it is well known that topological insulators can be split into those
without chiral (sublattice) symmetry and with chiral symmetry~\cite{Ryu2010}. It can then be shown
 that the edge of non-chiral disordered topological insulators are characterized by a band of delocalized state spanning the energy interval between the delocalized bulk states. They are delocalized along the edge while exponentially decaying into the bulk, as is expected from the proper edge states. At the same time, the edge of chiral topological insulators has at least one delocalized state at zero energy while the rest of the states may be localized by disorder. 

These statements generally match what is expected from the topological insulators from the study of sigma models or by using other methods. In particular, the localization of the one dimensional edge of two dimensional topological insulators is very well understood. There is no doubt, for example, that the edge of an integer Hall state is delocalized regardless of disorder, thanks to its chiral nature (absence of backscattering). Similar arguments can be made in case of other two dimensional topological insulators. 

The three dimensional topological insulators were also studied in recent years. Their two dimensional boundaries, in the presence of disorder, can be analyzed using sigma models. Those can belong to one of five symmetry classes. Of those, arguably the most important is the AII topological insulator, or the  standard strong time-reversal invariant topological insulators with spin-orbit coupling \cite{MooreBalents2007,FuKaneMele2007}. It has no sublattice symmetry. That insulator is known to have a fully delocalized edge, as confirmed in a variety of studies \cite{Nomura2007}. This agrees with our claim that all nonchiral insulators have a band of delocalized states at the boundary.

The remaining four insulators in 3D are all chiral. Among them, the simplest is the insulator with a sublattice symmetry only, known as AIII topological insulator. Its edge is described by two dimensional Dirac fermions with random gauge potential \cite{Ludwig1994}. Surprisingly, it was shown only a few years ago that  when the random gauge potential is zero on the average these insulators have either a fully delocalized edge or an edge with localized states with localization length which diverges as energy is taken to zero, depending on whether their topological invariant is even or odd \cite{Ostrovsky2007}. This results is obtained by mapping the problem at finite energy into a Pruisken-type sigma model with a topological term which corresponds to exactly the point of the integer quantum Hall transition it describes if the invariant is odd integer, and to the localized quantum Hall plateau if the invariant is even integer. However, this result is not completely robust: adding a constant magnetic field to the random gauge potential  shifts the coefficient of the Pruisken-type sigma model, potentially localizing all states except those at zero energy, in all cases \cite{Ostrovsky2014}.

The next is the insulator in class DIII represented by a superfluid $^3$He in its phase B. By mapping it into a sigma model, one finds that, similarly to the AIII case, this insulator has an edge which is either fully delocalized if the invariant is odd (like for $^3$He where it is exactly $1$) or with just one delocalized state at zero energy if the invariant is even. Technically this occurs because at finite energy an edge of such an insulator crosses over to the symmetry class AII which has a ${\mathbb Z}_2$ structure. 

The insulator in class CI (represented by an exotic spin-singlet superconductor \cite{Schnyder2009}) is known to have a fully localized edge except one state in the middle of the band, since at finite energy its edge crosses over to class AI, time-reversal invariant spin rotation invariant systems which were known for a very long time to be localized in two dimensions \cite{Abrahams1979}.

Finally, the topological insulator in class CII, the most exotic of the five topological insulators in three dimensions, has a fully localized edge except one state, as can be argued based on its mapping to the trivial (non topological) AII insulator at finite energy. 

All of these examples match what follows from the arguments which are presented in this paper. However, we note that a variety of chiral insulators, at least in three dimensions, have a fully delocalized edge going beyond the prediction of at least one delocalized state given here. Whether our method can be generalized to explain these additional features is not known. 

On the other hand, our method works not only in two or three dimensions, but  also for all topological insulators of  arbitrarily large spacial dimension where sigma models might be difficult to analyze. 

Finally we would like to point out that there exists an alternative method of studying topological insulators with disorder, based on the non-commutative Chern number and its generalizations \cite{Schulz2002,Schulz2003,Loring2010,Loring2011,Prodan2011}. These methods provide yet another way to look at the boundary of topological insulators with disorder \cite{Schulz2002}, which may well prove to be more powerful than the methods discussed here.

\section{Topological invariants of disordered insulators}

We start with non-chiral topological insulators in even dimensional space $d$. When disorder is absent they are characterized by the topological invariant
which can be constructed out of its Green's function \be \label{eq:green} G=\left[i \omega - {\cal H} \right]^{-1}.\ee  Assuming translational invariance, the Green's function takes a form of a matrix $G_{ab} (\omega, {\bf k})$ which depends on the frequency $\omega$ and the $d$-dimensional momentum ${\bf k}$ with indices $a$, $b$ labeling bands as well as spin and flavor, the invariant takes the form
\be \label{eq:top} N_d = C_d \, \epsilon_{a_0 \dots a_d} \tr \int d\omega d^d k \, G^{-1} \d_{a_0} G \dots G^{-1} \d_{a_d} G.
\ee
Here $C_d$ is a constant which makes $N_d$ an integer, known to be given by
\be C_d = - \frac{\left( \frac d 2 \right)!}{(2 \pi i)^{\frac d 2 + 1} \left( d+1 \right)!},
\ee
indices $\alpha_0$, $\alpha_1$, \dots, $\alpha_d$ run over $\omega$, $k_1$, \dots, $k_d$ (summation on indices always implied), $\epsilon$ is the Levi-Civita symbol and
$\tr$ is the trace over the matrix indices of $G_{ab}$. 
This invariant, if  Eq.~\rf{eq:green} is taken into account, is nothing but the Chern number of negative energy (filled) bands at $d=2$, second Chern number at $d=4$, etc.

Once the insulator is disordered, it is no longer translationally invariant and Eq. \rf{eq:top} loses any meaning. An alternative form for the topological invariant can be introduced in the following way. 

Following Ref.~[\onlinecite{Niu1985}] (see also Ref.~[\onlinecite{Essin2007}]), introduce the finite size system such that the wave function for each particle satisfies periodic boundary conditions with an additional phases $\theta_i$, where $i=1, \dots, d$ labels $d$ directions in space. In other words, \be \psi(x_i+L_i) = e^{i  \theta_i} \psi(x) \ee for a particular coordinate $x_i$ where $L_i$ is the size of the system
in the $i$-th direction. Then one introduces a Green's function $G_{\alpha \beta} (\theta)$ which is  no longer Fourier transformed, but which depends on the $d$ angles $\theta$. Here $\alpha$, $\beta$ label not only spin and flavor of the fermions but also the sites of the lattice.  The topological invariant is given by essentially exactly the same formula, \rf{eq:top}, 
but just interpreted in a slightly different way. The indices $\alpha_0$, $\dots$, $\alpha_d$ are now summed over $\omega$, $\theta_1, \dots, \theta_d$, and the symbol $\tr$ implies summation over all the matrix indices,
while the integral is performed over $d\omega d^d \theta$,
\be \label{eq:top1} N_d = C_d \, \epsilon_{a_0 \dots a_d} \tr \int d\omega d^d \theta \, G^{-1} \d_{a_0} G \dots G^{-1} \d_{a_d} G.
\ee
Here the integration over each $\theta$ extends from $-\pi$ to $\pi$ and the trace is over the matrix indices of $G_{\alpha \beta}$. 

In case when there is no disorder, the invariant introduced in this way coincides with the invariant defined in terms of momenta. Indeed, we can take advantage of translational invariance and reintroduce the momenta in \rfs{eq:top1}. Due to the periodic boundary conditions with the phases, 
 the momenta are restricted to the values (for each of the $d$ directions) $k^{(n)}_{i} = \left(2 \pi n_i +\theta_i \right)/L_i$ with $n_i$ being integers. 
 Integration over $\theta_i$ and summation over $n_i$ together are now equivalent to an unrestricted  integration over all the values of $k$, as in \rfs{eq:top}.
 Thus in the absence of disorder, Eqs.~\rf{eq:top} and \rf{eq:top1} are equivalent. 
 Yet unlike \rfs{eq:top}, the expression in terms of phases \rfs{eq:top1} can be used even in the presence of disorder when translational invariance is broken. 

If $d$ is odd, then \rfs{eq:top} is zero. Instead we follow Ref.~[\onlinecite{Ryu2010}] and consider insulators with chiral symmetry, such that there is a matrix $\Sigma$
\be \label{eq:sigma} \Sigma {\cal H} \Sigma = - {\cal H}.
\ee
As well known, only the  insulators with this symmetry have invariants of the integer type in odd spatial dimensions. The invariant itself, without disorder, can be written as follows
\be \label{eq:ci1} N_d = \frac{C_{d-1}}{2} \epsilon_{\alpha_1 \dots \alpha_d} \tr \int d^d k \, \Sigma \, V^{-1} \d_{\alpha_1} V  \dots V^{-1} \d_{\alpha_d} V.
\ee Here $\alpha_1, \dots \alpha_d$ are summed over $k_1$ to $k_d$ each, and \be V= \left. G \right|_{\omega=0}= -\cH^{-1}. \ee (We could equally well use $\cH$ instead of $V$
in the definition of the topological invariant, but use $V$ to smoothen the difference between even $d$ and odd $d$ cases). 
Again, in case if there is disorder present (which preserves symmetry \rfs{eq:sigma}, we can rewrite the invariant in terms of phases
\be \label{eq:ci2}  N_d = \frac{C_{d-1}}{2} \epsilon_{\alpha_1 \alpha_2 \dots \alpha_d} \tr \int d^d \theta \, \Sigma \, V^{-1} \d_{\alpha_1} V  \dots V^{-1} \d_{\alpha_d} V.
\ee
Here $\alpha_1, \dots, \alpha_d$ are summed over values  $\theta_1$ to $\theta_d$ each. 

Again, simple arguments can be given that these two expressions for $N_d$ in case when there is no disorder coincide. Once the disorder is switched on, \rfs{eq:ci1} loses any meaning, while \rfs{eq:ci2} can still be used. 

We will not separately study the invariants of the type ${\mathbb Z}_2$, as those can be obtained by dimensional reduction from the invariants of the type ${\mathbb Z}$ introduced above. This will be briefly discussed at the end of the paper. 

\section{Boundary of topological insulators} 

\subsection{Boundary of a disorder-free topological insulator}
Following our prior work \cite{Essin2011}, we would like to consider a situation where a domain wall is present such that the topological insulator is characterized by one value of the topological invariant  on the one side of the domain wall and another value on another side. We would like to examine the nature of the edge states forming at the boundary. 
\begin{figure}[htbp]
	\centering
	\includegraphics[width=0.3\textwidth]{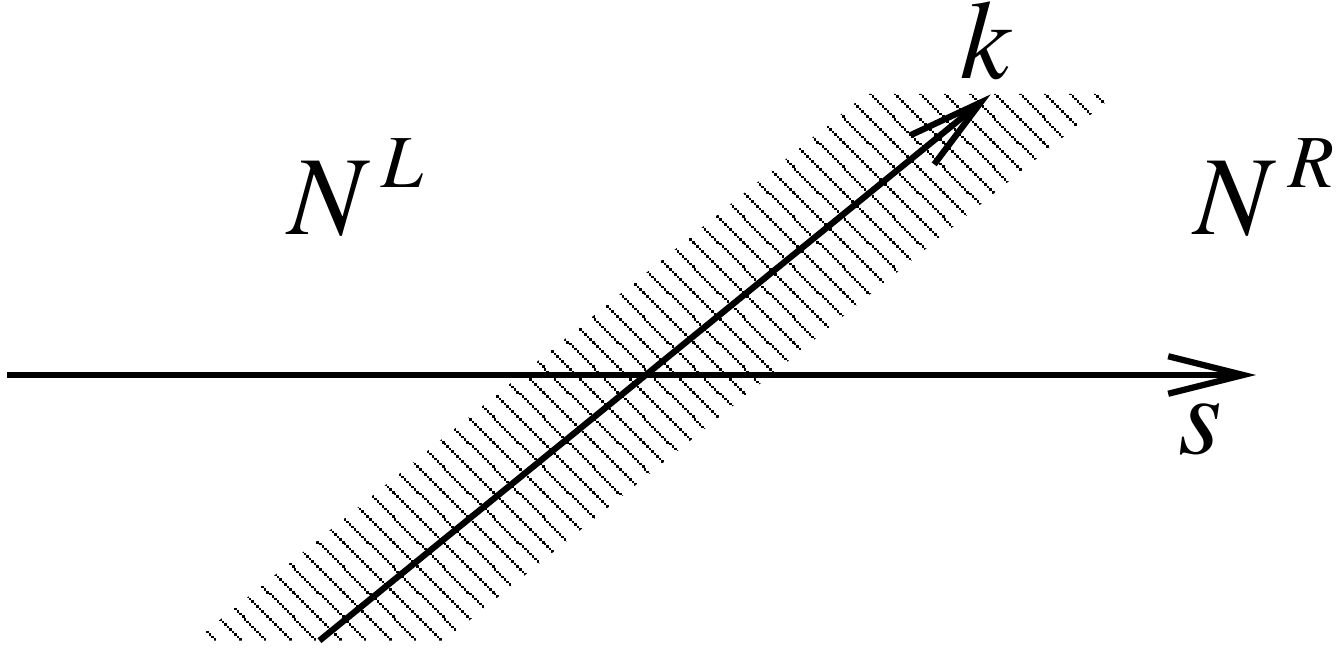}
	\caption{\label{boundary}
	A $d$-dimensional system with a domain wall separating two phases, one with a topological invariant $N^L$ at large negative coordinate $s$ and one with a topological
	invariant $N^R$ at large positive coordinate $s$. The insulator is translationally invariant in the direction perpendicular to $s$ so that direction can be spanned
	by the $d-1$ momenta $k$. }
\end{figure}

Let us first review the approach taken in Ref.~[\onlinecite{Essin2011}] in case when there is no disorder. 
It is possible to introduce the Green's function of the entire system $G_{ab}(\omega; k_1, \dots, k_{d-1}; s,s')$. Here $k_1$, $\dots$, $k_{d-1}$ span the boundary separating
two insulators, and $s$ is the coordinate perpendicular to the boundary where the system is not translationally invariant, see Fig.~\ref{boundary}. 
With its help it was furthermore possible to introduce the Wigner transformed Green's function
\begin{eqnarray} \label{eq:wig} && G_{ab}^W(\omega, {\bf k}, R) = \\ && \int dr \, e^{- i k_d r} \, G_{ab}\left(\omega; k_1, \dots, k_{d-1}; R + \frac r 2 , R - \frac r 2 \right). \nonumber
\end{eqnarray} Here ${\bf k}$ now denotes all $d$ momenta $k_1$, \dots, $k_d$. 
We can now introduce the concept of a Green's function on the far left of the boundary and the far right of the boundary,
\begin{eqnarray} G^R(\omega, {\bf k})  &=& \lim_{R \rightarrow \infty} G(\omega, {\bf k}, R), \cr
G^L(\omega, {\bf k})  &=& \lim_{R \rightarrow - \infty} G(\omega, {\bf k}, R).
\end{eqnarray}
These can now be used to calculate the topological invariant on the right and on the left of the boundary, $N^R_d$ and $N^L_d$ respectively, according to \rfs{eq:top} or \rfs{eq:ci1}, depending on whether $d$ is even or odd, with $G^R$ and $G^L$ substituted for $G$. Since we are considering a situation where the boundary separate two topologically distinct states, these two values are distinct, $N^R_d \not = N^L_d$. 

Furthermore, as it was discussed in Ref.~[\onlinecite{Essin2011}], there is also a boundary topological invariant, which can be defined with the help of the original
Green's function $G_{ab}(\omega; k_1, \dots, k_{d-1}; s, s')$  by
\begin{eqnarray}  \label{eq:boundaryinv} N^B_d &=& C_{d-2}\epsilon_{\alpha_0 \alpha_1 \dots \alpha_{d-1} } \int dS^{\alpha_0} X_{\alpha_1 \dots \alpha_{d-1}}, \cr
X &=& {\rm Tr} \,  \left[ G^{-1} \circ \d_{\alpha_1} G \circ  \dots \circ \d_{\alpha_{d-1}} G \right].
\end{eqnarray}
Here we introduced convenient notations, following Ref.~[\onlinecite{Essin2011}], where
\begin{eqnarray} A \circ B  &=& \sum_b \int ds' A_{ab}(s,s') B_{bc}(s',s''), \cr
{\rm Tr} \, A &=& \sum_a \int ds \, A_{aa}(s,s).
\end{eqnarray}
The integral in \rfs{eq:boundaryinv} is over a $d-1$-dimensional surface in the $d$-dimensional space formed by $\omega$, $k_1$, \dots, $k_{d-1}$. It was shown in Ref.~[\onlinecite{Essin2011}] that
\be \label{eq:bbr} N^R_d-N^L_d = N^B_d.
\ee (a much earlier work \cite{VolovikBook1} showed this for $d=2$). 

If the space is of odd dimensions, closely similar definition of  $N^B_d$ can be given, with \rfs{eq:bbr} still valid. For completeness, let us give them here. 
We now have a system with a chiral symmetry, implying that
\be \Sigma G(\omega) \Sigma = - G(-\omega).
\ee
The boundary invariant can now be defined with
\begin{eqnarray}  \label{eq:boundaryinvc} N^B_d &=& \frac{C_{d-3}}{2} \, \epsilon_{\alpha_1, \dots, \alpha_{d-1}} \int dS^{\alpha_1} \, X_{\alpha_2 \dots \alpha_{d-1}}, \\
X_{\alpha_2 \dots \alpha_{d-1}} &=& \, {\rm Tr} \, \left[ \Sigma V^{-1} \circ \d_{\alpha_2} V \circ \dots \circ V^{-1} \d_{\alpha_{d-1}} V \right]. \nonumber
\end{eqnarray}
Here \be V(k_1, k_2, \dots, k_{d-1}; s,s')  = \left. G(\omega, k_1, \dots, k_{d-1}; s,s') \right|_{\omega=0}.
\ee Indices $\alpha_1$ to $\alpha_{d-1}$ are summed over $k_1$, $k_2$, \dots, $k_{d-1}$. The integral is over a $d-2$ dimensional surface in the $d-1$ dimensional space formed by 
$k_1$, \dots, $k_{d-1}$. 
The bulk invariants can still be computed using \rfs{eq:ci1}, with $V^R = \left. G^R\right|_{\omega=0}$ substituted for $\cH$,  and similarly for $V^L$.

The boundary invariant $N^B_d$ can be useful in analyzing boundary theories of particular topological insulators. These boundary theories must have nonzero  boundary invariants, which in case when the boundary is between a non-trivial insulator and an empty space (whose invariant is zero) must be equal, up to a sign, to the bulk invariant. 

\subsection{Boundary with disorder}

We would like to generalize $N^B_d$ to the case when disorder is present. It is clear that we need to replace the momenta $k_1$, $\dots$, $k_{d-1}$ along the boundary with $d-1$ phases across the boundary $\theta_1, \dots, \theta_{d-1}$. However, it is not immediately clear what to do in the direction perpendicular to the boundary. 

\begin{figure}[htbp]
	\centering
	\includegraphics[width=0.45\textwidth]{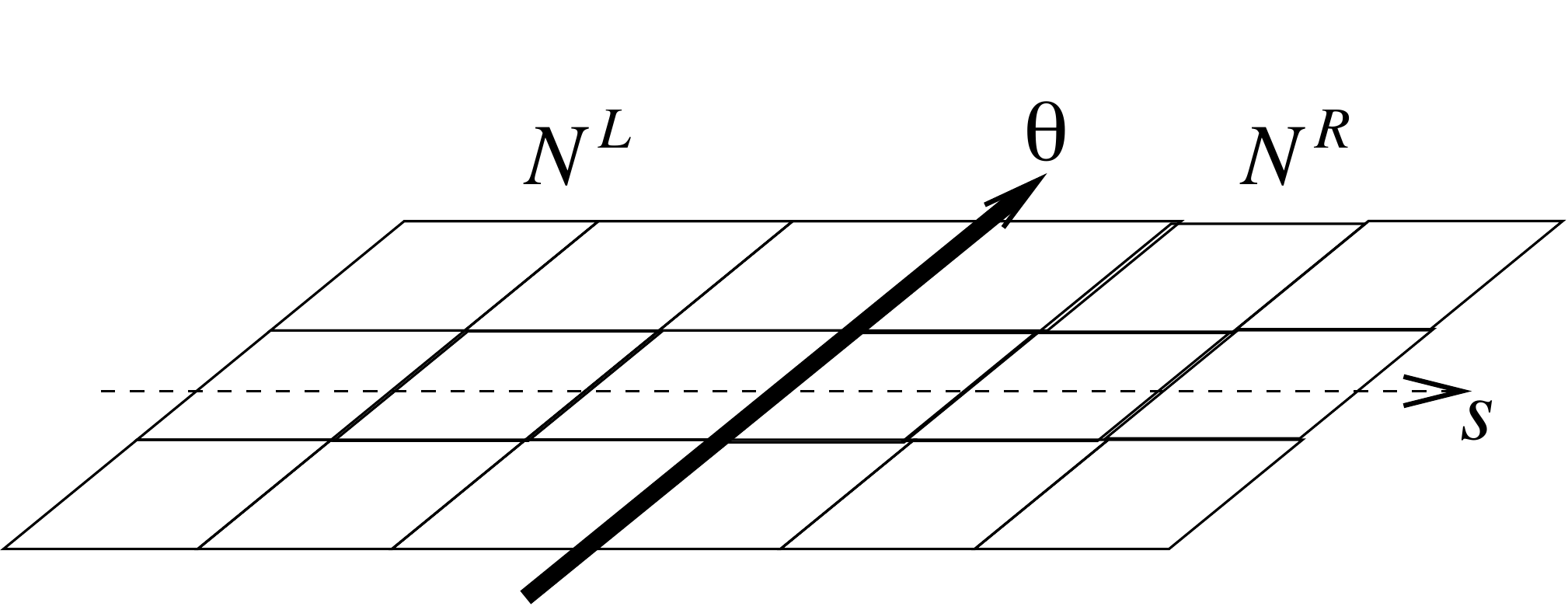}
	\caption{\label{disboundary}
	A $d$-dimensional system periodically repeated. Each rectangle represents a system with disorder. That disorder is identical in every  spatial replica of the system. However, a certain parameter in the Hamiltonian varies from replica to replica along the direction of the coordinate $s$ so that
	at some point the system has a boundary to a topologically distinct phase, indicated here by a think coordinate line, whose direction is spanned by the phases $\theta_1$ to $\theta_{d-1}$. }
\end{figure}

In order to deal with this direction, we use the following trick. Imposing phases across the system is tantamount to periodically replicating our system in space (with disorder being exactly the same in each replica), with the phase becoming equivalent to the  usual crystalline quasi-momentum. We will also periodically repeat the system in the direction perpendicular to the boundary. However, in that direction we should also ensure that the Hamiltonian changes close to the boundary, in such a way that once we move past  the boundary the system goes into a different topological class with a different invariant. We can make the Hamiltonian change slightly from a replica to a replica until the system goes through a transition between two adjacent replicas of the system where the boundary between two topologically distinct states resides. This is schematically shown in Fig.~\ref{disboundary}.  Labeling the replicas of the system by the (discrete) variable $s$, we can define the Green's function of the replicated system as
$G(\omega; \theta_1, \dots, \theta_{d-1}; s, s')$. Here $\theta_i$, $i=1, \dots, i=1$ phases are quasimomenta (or phases) across the $d-1$ dimensional boundary, and the remaining variables $s$, $s'$ label copies of the system in the direction perpendicular to the boundary. 

Given this Green's function we define the Wigner transformed function by Fourier series (compare with \rfs{eq:wig}
\begin{eqnarray} && G^W (\omega; \theta_1, \dots, \theta_d; S) = \cr && \sum_s e^{-i s \theta_d} G\left(\omega; \theta_1, \dots, \theta_{d-1}; S + \frac s 2, S - \frac  s 2 \right).
\end{eqnarray}
Here $S$ can be either integer or half-integer, and the summation over $s$ goes over either even integers or odd integers respectively. 
Given this function, we can again find the far left and far right Green's function
\begin{eqnarray} G^R(\omega; \theta_1, \dots, \theta_d)  &=& \lim_{S \rightarrow \infty} G(\omega; \theta_1, \dots, \theta_d, S), \cr
G^L(\omega; \theta_1, \dots, \theta_d)  &=& \lim_{S \rightarrow -\infty} G(\omega; \theta_1, \dots, \theta_d, S).
\end{eqnarray}
These can be used to define the left and right invariants according to \rfs{eq:top1} for even $d$, with $G^{L,R}$ substituted for $G$, or according to \rfs{eq:ci2} for odd $d$, with 
$\left. G^{L,R} \right|_{\omega=0}$ substituted for $V$.

Now the difference $N^R-N^L$ is again expected to be equal to $N^B$, the boundary invariant. This invariant is defined analogously with Eqs.~\rf{eq:boundaryinv} and \rf{eq:boundaryinvc} for $d$ even and odd respectively, with the momenta replaced by the phases. If $d$ is even, 
\begin{eqnarray}  \label{eq:boundaryinvphases} N^B_d &=& C_{d-2}\epsilon_{\alpha_0 \alpha_1 \dots \alpha_{d-1} } \int dS^{\alpha_0} X_{\alpha_1 \dots \alpha_{d-1}}, \cr
X &=& {\rm Tr} \,  \left[ G^{-1} \circ \d_{\alpha_1} G \circ  \dots \circ \d_{\alpha_{d-1}} G \right].
\end{eqnarray}
where the indices $\alpha_i$ are summed over $\omega$, and $\theta_1$, \dots, $\theta_{d-1}$, and the integration over $s$ (implicit in the definitions of Tr and $A \circ B$) is replaced by summation. 

Similarly for $d$ odd we can use construct the boundary invariant by using 
\begin{eqnarray}  \label{eq:boundaryinvcphases} N^B_d &=& \frac{C_{d-3}}{2} \, \epsilon_{\alpha_1, \dots, \alpha_{d-1}} \int dS^{\alpha_1} \, X_{\alpha_2 \dots \alpha_{d-1}}, \\
X_{\alpha_2 \dots \alpha_{d-1}} &=& \, {\rm Tr} \, \left[ \Sigma V^{-1} \circ \d_{\alpha_2} V \circ \dots \circ V^{-1} \d_{\alpha_{d-1}} V \right], \nonumber
\end{eqnarray}
where $V= \left. G \right|_{\omega=0}$, and $\alpha_i$ are summed over $\theta_1$, \dots, $\theta_{d-1}$.

The integrals in Eqs.~\rf{eq:boundaryinvphases} and \rf{eq:boundaryinvcphases} are taken over closed surfaces in the $d$ dimensional (if $d$ even) or $d-1$ dimensional (if $d$ is odd) space. 
The derivation of \rfs{eq:bbr} for this disordered case is essentially the same as in the disorderless case presented in Ref~[\onlinecite{Essin2011}]. The only (minor) difference is that here $S$ and $s$ are discrete variables. But since we construct our boundary in such a way that the system changes very little as $S$ is increased by $1$, it can essentially be treated as a continuous variable, and the discussion of Ref.~[\onlinecite{Essin2011}] still applies. 

\subsection{Analysis of the boundary invariants in the presence of disorder}

We would now like to analyze the the boundary topological invariants. Take the invariant \rfs{eq:boundaryinvphases} (applicable when $d$ is even). The integral in \rfs{eq:boundaryinvphases} is computed over a closed surface in the $d$ dimensional space formed by the frequency $\omega$ and $d-1$ phases spanning the boundary of the system. The choice of the surface is arbitrary, as long as it encloses the points or surfaces where $G$ is singular (see Ref.~[\onlinecite{Essin2011}) for the discussion concerning the  surface choice).

It is then natural to choose  as a surface to be integrated over the two planes at two values of the  phase $\theta_{d-1} = \pm \Lambda$, as as it is done in our prior work \cite{Essin2011} (the choice of $\theta_{d-1}$ is arbitrary; any of $\theta_i$ can be chosen for this construction). Here $\Lambda$ is such that all the singularities of $G$
lie between these two planes in the $\theta$-space. Then the boundary invariant can be rewritten as essentially a bulk invariant in the $d-1$-dimensional space formed by frequency and $d-2$ phases across the boundary, $\theta_1, \dots, \theta_{d-2}$, with 
$\theta_{d-1}$ fixed. More precisely, we can define
\begin{eqnarray} N_{d-2}(\theta_{d-1}) &=& C_{d-2}  \, \epsilon_{\alpha_0 \dots \alpha_{d-2}}  \int d\omega d^{d-2} \theta \, X, \\ 
X &=& {\rm Tr} \, \left[ G^{-1} \circ \d_{\alpha_0} G \circ \dots G^{-1} \circ \d_{\alpha_{d-2}} G \right]. \nonumber
\end{eqnarray}
Here $\alpha_0$, $\dots$, $\alpha_{d-2}$ are summed over $\omega$, $\theta_1$, $\dots$, $\theta_{d-2}$. 

Then we can rewrite $N^B_d$ as
\be \label{eq:sur} N_d^B = N_{d-2}  (\Lambda) - N_{d-2}(-\Lambda).
\ee
Therefore, if $N^B_d$ is not zero, $N_{d-2}$ is a function of $\theta_{d-1}$ in such a way that as $\theta_{d-1}$ changes from $-\Lambda$ to $\Lambda$, $N_{d-2}$ changes by an amount equal to $N^B_d$. And as we discussed if the boundary we study is the boundary of a topological insulator with nonzero bulk invariant $N_d$, $N^B_d=N_d \not =0$. 
 
Now $N_{d-2}$ is a topological invariant. The only way for it to change as a function of $\theta_{d-1}$ is if there is some special value $-\Lambda<\theta_{d-1}^c<\Lambda$ such that at $\theta_{d-1}$ equal to this value, $G$ becomes singular. Then at that value $N_{d-2}$ is not well defined, and the difference $N_{d-2}(\Lambda) - N_{d-2}(-\Lambda)$ can be nonzero. 

$G$ is related to the Hamiltonian by $G= \left[ i \omega - \cH \right]^{-1}$. The only way for it to be singular if $\omega=0$ and $\cH$ has a zero eigenvalue. That means, there is a single particle energy level at the boundary whose energy is zero at $\theta_{d-1}= \theta^c_{d-1}$. At the same time, when $\theta_{d-1}=\pm \Lambda$, that energy level is not zero.

We can now invoke a well known criterion of localization which states that a localized level's energy cannot shift as a function of the phase imposed across a disordered system. Therefore, in order for $N^B_d$ to be nonzero, there has to be at least one energy level whose energy depends on $\theta_{d-1}$. That level must be delocalized. 

Furthermore, the system we study must have more than one localized energy level. Indeed, zero was an arbitrary reference point for the energy. We can always consider a Hamiltonian shifted by some chemical potential $\mu$ (a position-independent constant),
\be \cH' = \cH - \mu.
\ee We can repeat all the arguments for this shifted Hamiltonian. As long as this shift does not change $N_d$ (the bulk invariant), there should be a delocalized state at new zero energy, or at energy $\mu$ of the original model. Now we can anticipate that the bulk system has states at all energies, but states in the energy interval $-\Delta$ to $\Delta$ are all localized. This concept of $\Delta$ generalizes the concept of a gap in case of a disorder-free system. Then for any
 $-\Delta<\mu<\Delta$, the bulk invariant $N_d$ is insensitive to $\mu$.  
Then  the system has a delocalized state at any energy $\mu$ which spans the interval $-\Delta$ to $\Delta$. This concludes the argument about the delocalized states at the edge of any even-dimensional insulator. 

The situation with odd-dimensional insulators is somewhat more restrictive. Their boundary invariant \rfs{eq:boundaryinvcphases} can still be rewritten in a way equivalent to \rfs{eq:sur}, with $N_{d-2}(\theta_{d-1})$ given by
\begin{eqnarray} N_{d-2} (\theta_{d-1}) &=& \frac{C_{d-3}}2 \epsilon_{\alpha_1 \dots \alpha_{d-2}} \int d^{d-2} \theta \, X, \\ \nonumber
X &=& {\rm Tr} \left[ \Sigma V^{-1} \circ \d_{\alpha_1} V \circ \dots \circ V^{-1} \d_{\alpha_{d-2}} V \right].
\end{eqnarray}
where as before $V = \left. G \right|_{\omega=0}$. Just as before, at the boundary of a topologic insulator $N_d = N^B_d = N_{d-2}(\Lambda)-N_{d-2}(-\Lambda) \not =0$. It again follows that there is an energy level which crosses zero as a function of $\theta_{d-1}$ at some value $\theta_{d-1}^c$. That level must be delocalized. 

However, no other delocalized levels can be generally expected. Indeed, the Hamiltonian cannot be shifted by a chemical potential, as before, because we must ensure the symmetry \rfs{eq:sigma}. An arbitrary  chemical potential added to the Hamiltonian  breaks it. Therefore, we expect only one delocalized state close to zero energy (in the limit of the infinitely large system, exactly at zero energy). All other states are generally localized. 

This is indeed what is expected from the chiral systems as explained in the introduction. Some of the chiral systems have only one delocalized state at zero energy. Yet others have many delocalized states spanning some energy interval centered around zero, similarly to the non-chiral disordered insulators. The arguments given here cannot establish which of the chiral systems will have a fully delocalized edge. All one can establish is that at least one state at zero energy must be delocalized. 

\subsection{Edge of a topological insulator with ${\mathbb Z}_2$ invariant}

Finally, let us address the rest of topological insulators described by a invariant ${\mathbb Z}_2$   taking just two values, $0$ and $1$. All topological insulators are described by either an integer invariant of the types discussed here earlier or the invariant of the type ${\mathbb Z}_2$ as is well summarized in Ref.~\onlinecite{Ryu2010}.

 Systems with   ${\mathbb Z}_2$ topological invariant can be understood as a dimensional reduction of the system with an invariant described in this paper by residing in higher dimensions \cite{Qi2008,Ryu2010}. One can imagine that higher dimensional system having disorder which does not vary in the spatial direction we plan to eliminate. Those dimensions can be spanned by momenta in the Green's functions. Setting those momenta to zero we obtain the dimensionally reduced system with the invariant ${\mathbb Z}_2$. The boundary states of both higher dimensional ${\mathbb Z}$ ``parent" and lower dimensional ${\mathbb Z}_2$ ``descendant" system must be delocalized in the same way. Thus we find that the boundary of ${\mathbb Z}_2$ topological insulators have a fully delocalized band if they do not have chiral symmetry or  at least a single delocalized state at zero energy if they do have chiral symmetry.

\section{Conclusions}

We examined the boundary states of disordered topological insulators and were able to understand their localization properties by directly examining the topological invariants of these disordered systems. In doing so, we reproduced what should be considered a widely anticipated answer. Nevertheless it was only for two and some three dimensional insulators that these answer has been derived. Further elaborations were based on the approach of  the sigma models with WZW-type terms, and on their mostly conjectured behavior (although in some low dimensional cases this can be derived). Here we derived this answer without any conjectures in an arbitrary number of dimensions.

Finally in view of the existence of other methods to look at boundaries of topological insulators \cite{Prodan2011}, 
it would be interesting to further explore 
the connection between these methods and the one discussed here and see if this could shed additional light on the structure of the boundary states in disordered topological insulators.

\acknowledgements
The authors are grateful to P. Ostrovsky for sharing insights concerning the boundaries of three dimensional disordered topological insulators. This work was supported by the NSF grants DMR-1205303 and PHY-1211914 (VG), and by the Institute for Quantum Information and 
Matter, an NSF Physics Frontiers Center with support of the Gordon and 
Betty Moore Foundation through Grant GBMF1250 (AE).

\bibliography{shortbib}

\end{document}